\documentclass[aps,prl,twocolumn,superscriptaddress,floatfix]{revtex4-2}
\usepackage{amsmath}
\usepackage{graphicx}
\usepackage{color}

\begin{document}

\title{Interfacial Oxidation Enables Charge-Transfer Contacts and Degenerate n-Doping in Monolayer MoS$_2$}


\author{Marco Bianchi}
\affiliation{Elettra - Sincrotrone Trieste S.C.p.A, Trieste, Italy}
\author{Daniel Lizzit}
\affiliation{University of Udine, Udine, Italy}
\author{Alberto Turoldo}
\affiliation{University of Trieste, Trieste, Italy}
\author{Ezequiel Tosi}
\affiliation{Instituto de Ciencia de Materiales de Madrid (ICMM - CSIC), Madrid, Spain}
\author{Paolo Lacovig}
\affiliation{Elettra - Sincrotrone Trieste S.C.p.A, Trieste, Italy}
\author{Monika Schied}
\affiliation{CNR - Istituto Officina dei Materiali (IOM), Trieste, Italy}
\author{Davide Curcio}
\affiliation{CNR - Istituto Officina dei Materiali (IOM), Trieste, Italy}
\author{Charlotte E. Sanders}
\affiliation{ArtemisProgram,U.K.Central Laser Facility, STFC Rutherford Appleton Laboratory, Didcot,UnitedKingdom}
\affiliation{School of Physics and Astronomy, University of St Andrews, St Andrews, United Kingdom}
\author{Silvano Lizzit}
\affiliation{Elettra - Sincrotrone Trieste S.C.p.A, Trieste, Italy}
\email{silvano.lizzit@elettra.eu}
\author{Philip Hofmann}
\affiliation{Department of Physics and Astronomy, Aarhus University, Aarhus, Denmark}
\date{\today}

\begin{abstract}
High contact resistance remains a central obstacle to the integration of two-dimensional (2D) semiconductors in electronic devices. Recent advances have demonstrated that contact performance can be dramatically improved through interface engineering, including the use of group-V semimetals and charge-transfer contacts based on strong interfacial doping. Here, we show that controlled interfacial oxidation provides an effective route to convert a semimetal contact into a charge-transfer contact that degenerately $n$-dopes single layer MoS$_2$. Using a combination of angle-resolved photoemission spectroscopy, X-ray photoelectron diffraction, low-energy electron diffraction and scanning tunnelling spectroscopy, we demonstrate that putting single layer MoS$_2$ in contact with a pristine Bi layer merely results in weak doping, whereas oxidation of the Bi layer leads to a pronounced occupation of the MoS$_2$ conduction band with an electron density on the order of $10^{13}$~cm$^{-2}$. The cause of this strong electron doping is the fact that an ultrathin $\beta$-Bi$_2$O$_3$ layer forms below the MoS$_2$ and that this has a particularly low work function, thereby acting as an efficient electron donor to MoS$_2$. Interfacial oxidation thus emerges as a powerful design knob for engineering charge-transfer contacts to 2D semiconductors.
\end{abstract}

\maketitle

\section{Introduction}\label{sec1}

Single-layer (SL) transition metal dichalcogenides (TMDCs) such as SL MoS$_2$ have a high potential for future (opto)electronic applications, including field-effect transistors \cite{Lembke:2015aa}, spin and valley physics \cite{Schaibley:2016aa}, superconductivity \cite{Saito:2015aa}, and excitonic devices \cite{Mueller:2018aa}. A central obstacle to their technological implementation, however, remains the realization of low-resistance and reproducible electrical contacts. 
The formation of ohmic contacts to SL TMDCs is challenged by several factors: the chemical sensitivity of the monolayer during device processing \cite{Allain:2015aa}, the formation of sizable Schottky barriers at conventional metal interfaces \cite{Das:2013ab,Allain:2015aa}, and the prevalence of interface disorder and metal-induced gap states that give rise to Fermi-level pinning \cite{Kang:2014aa,Farmanbar:2015aa}.
 As a result, establishing high-quality contacts to 2D semiconductors has become a major research focus, and alternative strategies such as van der Waals (vdW) contacts have been explored to achieve atomically abrupt and damage-free interfaces \cite{Ma:2024aa}.

Substantial recent progress has been achieved by employing the group-V semimetals bismuth and antimony as contact materials, leading to exceptionally low contact resistances and even approaching the quantum limit \cite{Shen:2021vx,Li:2023ab}. The underlying rationale is that the low density of states near the Fermi level in these semimetals may facilitate a favorable band alignment with the conduction band of SL MoS$_2$, potentially suppressing Schottky barriers. Indeed, transport experiments on Bi contacts have suggested strong electron doping and metallic behavior of MoS$_2$ in the contact region, supported by density functional theory calculations \cite{Shen:2021vx}. In addition, orientation-dependent hybridization between semimetal and TMDC states has been proposed as a further route to enhance carrier injection \cite{Li:2023ab}, analogous to the strong hybridization observed for SL WS$_2$ on Ag(111) \cite{Dendzik:2017ab}.

An alternative and conceptually distinct approach is provided by so-called charge-transfer contacts, in which the 2D semiconductor is doped by an adjacent electron donor or acceptor layer rather than by direct chemical substitution. A prominent example is the use of $\alpha$-RuCl$_3$ as an electron acceptor to achieve strong $p$-type doping of TMDCs, combined with graphene as an electrical contact \cite{Pack:2024aa}. 
More generally, such remote or modulation doping strategies—where charge transfer occurs without direct incorporation of dopants into the 2D lattice—have been proposed as a means to tune carrier densities while preserving the structural integrity of the 2D material \cite{Arora:2025aa,Wang:2025aa,Tang:2025aa}. Despite these advances, the microscopic electronic structure at semimetal--TMDC interfaces, and in particular the role of interfacial chemistry in enabling or suppressing charge transfer, remains poorly understood.

In this context, direct experimental access to the band alignment and carrier distribution at the contact interface is highly desirable. Angle-resolved photoemission spectroscopy (ARPES) has provided detailed insight into the electronic structure of epitaxial TMDC monolayers on noble metal substrates, where atomically well-defined interfaces can be achieved \cite{Miwa:2015aa,Sanders:2016aa,Arnold:2018ab,Stan:2019aa}. However, in the case of semimetal contacts such as Bi and Sb, epitaxial growth of SL MoS$_2$ has so far not been realized, precluding a similarly direct spectroscopic characterization of these technologically relevant interfaces.

In the present work, we address this limitation by combining epitaxial growth and intercalation to realize a well-defined semimetal--TMDC interface that is directly accessible to spectroscopic characterization. Specifically, we first grow epitaxial SL MoS$_2$ on Au(111) and subsequently intercalate Bi beneath the monolayer, thereby forming a contaminant-free MoS$_2$/Bi interface while preserving access to the MoS$_2$ band structure, band alignment and carrier concentration by ARPES. 

We find that the intercalation of a pristine Bi layer leads to a modest increase in the $n$-type doping of SL MoS$_2$ relative to the Au(111) substrate, but does not result in a detectable occupation of the MoS$_2$ conduction band minimum. This is in marked contrast to expectations based on transport experiments and theoretical studies of Bi and Sb contacts, which have suggested strong electron doping and metallization of MoS$_2$ in the contact region \cite{Lee:2015wm,Shen:2021vx,Li:2023ab}. 

Strikingly, controlled oxidation of the intercalated Bi layer beneath SL MoS$_2$ \cite{Lizzit:2012aa} qualitatively changes the electronic response. Upon oxidation, we observe a pronounced occupation of the MoS$_2$ conduction band, corresponding to an electron density  of $1.6 \times 10^{13}$~cm$^{-2}$, with the Fermi level located within the MoS$_2$ conduction band. 
This level of electron accumulation is comparable in magnitude to the hole densities achieved in recently reported charge-transfer contacts based on $\alpha$-RuCl$_3$ \cite{Pack:2024aa}, but with opposite carrier polarity. The oxidized Bi interlayer thus realizes a degenerate \emph{n}-type charge-transfer contact to MoS$_2$, effectively rendering the MoS$_2$ layer metallic in the contact region. These results demonstrate that interfacial oxidation can convert a semimetal contact into an efficient charge-transfer contact, establishing interfacial chemistry as a powerful engineering variable for tailoring low resistance contacts.

\section{Results}\label{sec2}

\begin{figure*}
\includegraphics[width=1.0\textwidth]{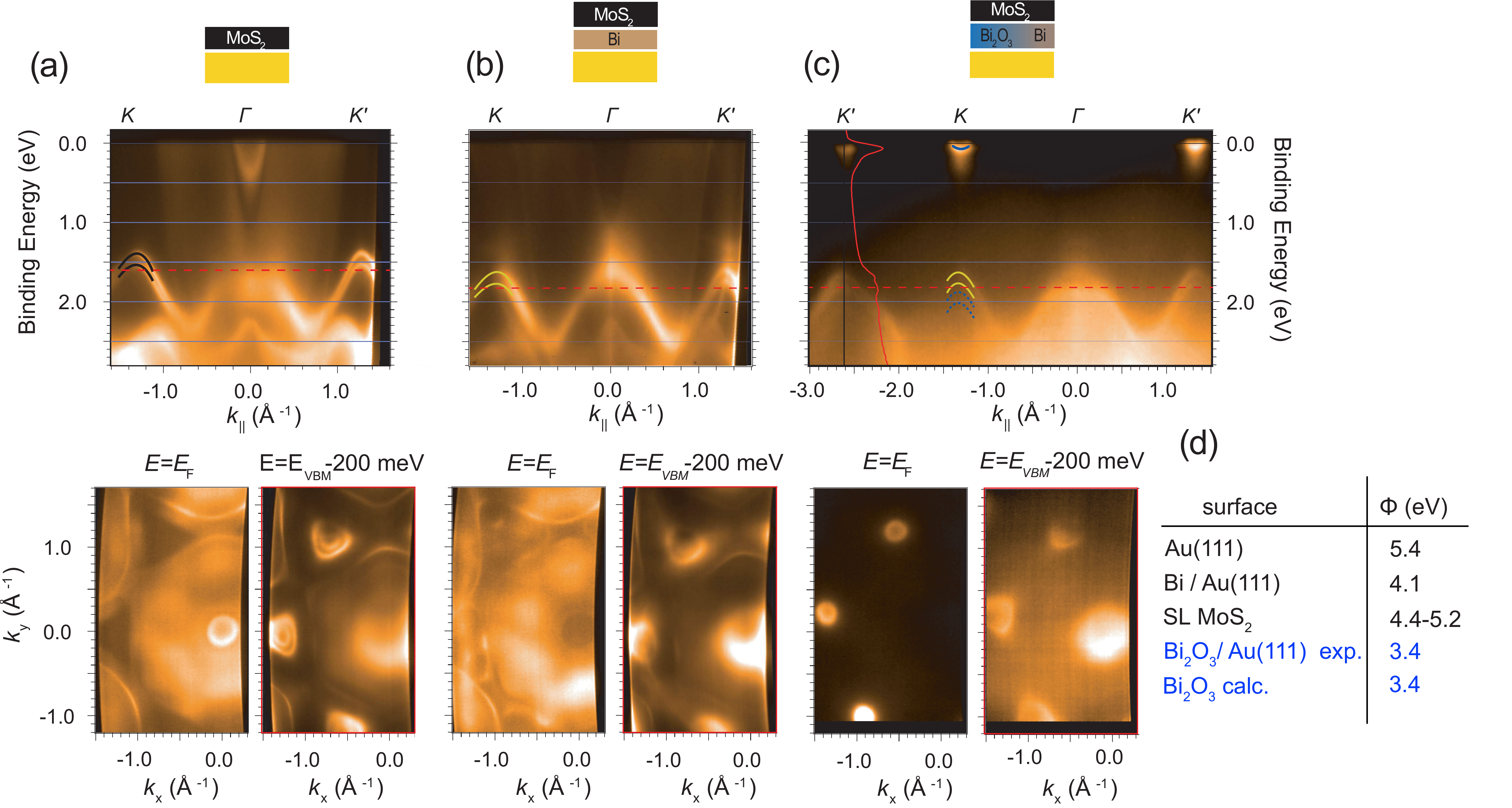} 
  \caption{(Color online) Photoemission intensity along the $\mathrm{K}-{\Gamma}-\mathrm{K}^{\prime}$ direction (upper panels) and momentum-dependent photoemission intensity at the Fermi level and 200~meV below the SL MoS$_2$ valence band maximum, as marked by the dashed red lines in the upper panels (lower panels).  (a)  As-grown  epitaxial single layer MoS$_2$ on Au(111).  (b) Bi intercalated SL MoS$_2$. (c) SL MoS$_2$ on bismuth oxide on Au(111). Black, yellow and blue curved lines indicate the dispersion near the SL MoS$_2$ valence band maxima for the clean, Bi-intercalated and Bi oxide substrate, respectively. The red line in panel (c), upper row, is an energy distribution curve taken at the $\mathrm{K'}$ point, along the vertical black line in order to estimate the position of the valence band maximum for the Bi oxide case. (d) Table with the measured work functions $\Phi$ for the indicated systems and the calculated work function for Bi-terminated $\beta$-Bi$_2$O$_3$. The value range for SL MoS$_2$ is from Refs. \cite{Robinson:2015aa,Choi:2014aa}.} 
  \label{fig:ARPES}
\end{figure*}

The substrate-dependent doping in SL MoS$_2$ and especially the surprising highly degenerate doping when placing the material on bismuth oxide is directly tracked by ARPES in Fig.~\ref{fig:ARPES}. The top panel gives the dispersion of the states along the $\mathrm{K}$-${\Gamma}$-$\mathrm{K}^{\prime}$ line of the Brillouin zone, clearly showing the spin-split SL MoS$_2$ valence band maximum (VBM) at  $\mathrm{K}$ (the corresponding bands are overlayed by black lines close to the left $\mathrm{K}$ point). The VBM states show minimal interaction with the Au(111) states due to their location in a projected bulk band gap \cite{Takeuchi:1991aa}. The local valence band maximum at ${\Gamma}$ is not clearly visible due to hybridisation with the Au(111) states. 
Under each of the panels showing dispersions in  (a)-(c) are constant energy contours corresponding, in each case, to the Fermi energy $E_\mathrm{F}$ (left side) and to an energy 200~meV below the SL MoS$_2$ VBM (right side). In panel (a), 
 SL MoS$_2$ does not contribute to any states at $E_\mathrm{F}$ due to its semiconducting nature. The constant energy contour 200~meV below the VBM, on the other hand, reveals the characteristic triangular contours around the $\mathrm{K}$ points, split by the spin-orbit interaction \cite{Miwa:2015ab}. Note that the in-plane lattice constant of Au(111) and SL MoS$_2$ is not the same and thus the Au(111) $\overline{\mathrm{K}}$ point is only approximately equal to the SL MoS$_2$ K point. The difference in lattice constants leads to a moir\'e superstructure that is clearly visible in scanning tunnelling microcopy (STM) and low energy electron diffraction (LEED) (see Appendix) but the moir\'e-induced scattering is not sufficiently strong to create observable replicas of the SL MoS$_2$ band structure in ARPES. These results are in excellent agreement with the literature \cite{Miwa:2015ab,Bruix:2016aa,Bana:2018aa}.

Upon Bi deposition at 373~K, new SL MoS$_2$ bands gradually appear at 250~meV higher binding energy, at the expense of the original bands (see Appendix). As more Bi is deposited, the original bands eventually disappear and only the high binding energy set of bands remains, as seen in  Fig.~\ref{fig:ARPES}(b).
This can be explained by the gradual intercalation of Bi under the epitaxial SL MoS$_2$, as will be demonstrated below, combined with the different band alignment at a SL MoS$_2$ interface with gold and bismuth. Interestingly, Bi intercalation merely leads to rigid shift of the bands without indications of strong SL MoS$_2$-Bi hybridisation. If anything, the local valence band maximum at ${\Gamma}$ appears to be more distinct than for SL MoS$_2$ on Au(111), suggesting that SL MoS$_2$ on Bi is more free-standing than on Au. 

Even after full Bi intercalation, there is no indication of a SL MoS$_2$ conduction band minimum occupation near $\mathrm{K}$ in the photoemission intensity at $E_\mathrm{F}$, despite the clearly increased $n$-doping and the fact that the VBM is now at a binding energy of more than $\approx$1.6~eV, approaching (but not exceeding) the expected band gap size of this system, which is between 1.7 and 2.0~eV \cite{Antonija-Grubisic-Cabo:2015aa,Bruix:2016aa}. These results are in contrast to the theoretical findings in Refs. \cite{Lee:2015wm,Shen:2021vx} which place the MoS$_2$ conduction band minimum $\approx$50~meV below $E_\mathrm{F}$ when putting the layer in contact with Bi.  However, such a comparison needs to be done with some care. First of all, there is a structural difference between the Bi layer on Au(111) and the (111) oriented Bi slab assumed in the calculations, such that the substrate electronic structure could be  different. Moreover, the slab thickness in the calculations are either bilayer Bi \cite{Lee:2015wm} or a 20~nm thick (111) oriented Bi film \cite{Shen:2021vx}, neither of which is a good approximation for the bulk surface of a (111) crystal due to the low effective masses of the bulk carriers and the resulting very long length scale for quantum confinement in Bi films \cite{Ogrin:1966aa}.

Surprisingly, a strongly degenerate doping of the SL MoS$_2$ emerges upon oxidising the intercalated Bi layer. This is shown in Fig.~\ref{fig:ARPES}(c). Now the conduction band minima are clearly visible at the $\mathrm{K}$ points both in the dispersion and in the photoemission intensity at $E_\mathrm{F}$. Indeed, the occupation is sufficient for the conduction band to appear as distinct circular contours in the latter, allowing us to estimate the electron density as $1.6~\times~10^{13}$~cm$^{-2}$.
Again, the doping appears consistent with a rigid shift of the Fermi level. The VBM at K has moved  down in energy by another 250~meV and is now only barely visible (marked by blue lines) below the remaining bands of the not-oxidised regions (marked by yellow lines). Its energy position can be estimated from a fit to the photoemission intensity along the black line at the  left  $\mathrm{K'}$ point of Fig.~\ref{fig:ARPES}(c), shown by a red solid line.   

The emergence of degenerate $n$-type doping in SL MoS$_2$ on the oxidized Bi film can be rationalized by the pronounced reduction of the substrate work function upon oxidation. To see this, we report in Fig.~\ref{fig:ARPES}(d)  work function measurements for the separately prepared clean surfaces -- without the top SL MoS$_2$ (see Appendix for details of the work function measurements). For clean Au(111), we find a work function of 5.4~eV, in excellent agreement with the literature \cite{Huber:1966aa}.  This is lowered to the expected value of 4.1~eV \cite{Michaelson:1977}  when the surface is covered with a Bi film corresponding to the intercalated amount in  Fig.~\ref{fig:ARPES}(b). Proceeding with the oxidation of this film transforms the layer into a phase that we identify as $\beta$-Bi$_2$O$_3$ (see below) with an estimated thickness of $\approx$25~\AA. This drastically reduces the work function to 3.4~eV, in excellent agreement with density functional theory (DFT) calculations for a thin  $\beta$-Bi$_2$O$_3$  film, also reported in Fig.~\ref{fig:ARPES}(d)  (for details of the calculations, see Appendix). Given that the measured work function for SL MoS$_2$ is between 4.4~eV and 5.2~eV~\cite{Robinson:2015aa,Choi:2014aa}, the work function changes of the substrate explain the observed trend of an increased electron transfer to the SL MoS$_2$ that is eventually strong enough to cause the degenerate $n$ doping. 

Charge transfer between SL MoS$_2$ and underlying substrates has been reported previously for mechanically exfoliated layers on a variety of metals and oxides, where substrate-induced $n$-type doping is commonly observed and, in some cases, even reaches the degenerate regime \cite{Rai:2015aa,Leonhardt:2019aa,McClellan:2021aa}. In these systems, the doping is typically attributed to fixed charges, oxygen vacancies, or interface states in the substrate, and is often regarded as an extrinsic or parasitic effect. In contrast, the present results demonstrate a contact-mediated charge-transfer mechanism in which the electronic response of SL MoS$_2$ can be deliberately switched from non-degenerate to degenerately $n$-doped by controlled interfacial oxidation of a semimetal contact layer.
 
\begin{figure*}
    \includegraphics[width=1.0\textwidth]{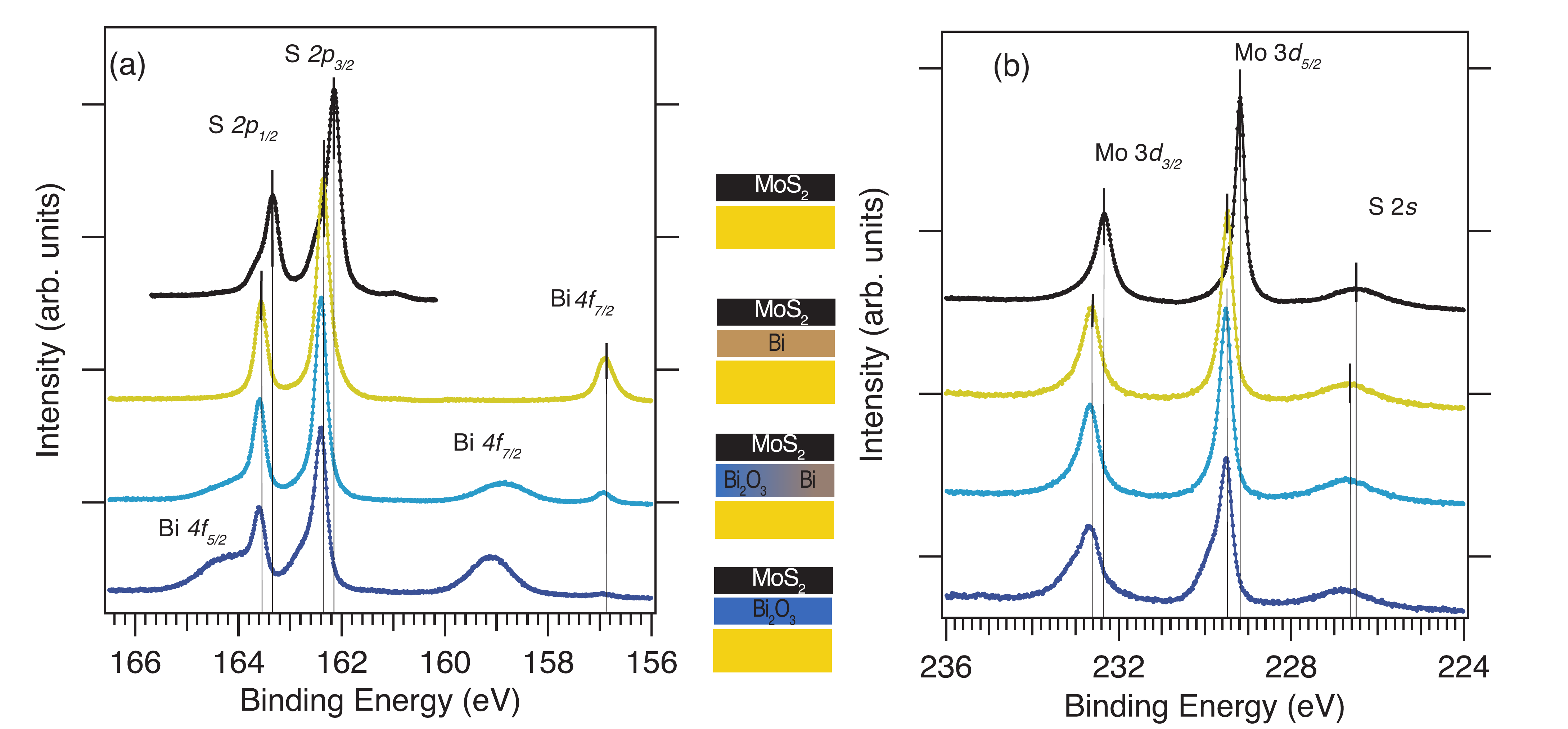}
    \caption{Bi intercalation and oxidation tracked by XPS on the (a) S $2p$ and  Bi $4f$  core levels at  h$\nu=260$~eV and (b) Mo $3d$ and  S $2s$  core levels at  h$\nu=360$~eV: Pristine SL MoS$_2$  on Au (black line); Bi intercalated (yellow line); oxidized system for different oxidation steps: in vacuum and in air (light and dark blue respectively).}\label{fig:XPS_S2p}
\end{figure*}

Structural and chemical information about the intercalation and the oxide formation can be gained by X-ray photoemission spectroscopy (XPS) on the S $2p$, Mo $3d$ and Bi $4f$ core levels. These results are shown in Fig. \ref{fig:XPS_S2p} for each step of the preparation. Starting with S $2p$ for pristine SL MoS$_2$ on Au(111) in Fig. \ref{fig:XPS_S2p}(a), each of the spin-orbit split components shows an additional splitting of about 290~meV. These two components have been identified as stemming from the ``top'' and ``bottom'' sulphur atoms of the MoS$_2$  \cite{Bana:2018aa}. The splitting arises because of their different chemical environment in which the top sulphur is exposed to vacuum and the bottom sulphur binds to the Au(111) surface.  The bottom component is here seen as a high binding energy shoulder with a lower intensity due to the attenuation of the photoelectrons in the MoS$_2$ layer. The splitting between bottom and top components is lost upon Bi intercalation, as seen by the narrower and more symmetric S $2p$ peaks. This is ascribed to a weaker interaction between sulphur and Bi, leading to a more free-standing nature of the SL MoS$_2$ layer as a consequence. These results are consistent with the ARPES observation of a more distinct valence band dispersion around ${\Gamma}$ as the hybridisation between the MoS$_2$ valence band and the substrate is reduced. 
Bi intercalation leads to an overall shift of the S $2p$ peaks to higher binding energy. This is consistent with the overall increased $n$-doping seen in the valence band by ARPES in Fig. \ref{fig:ARPES}(b). Finally, oxygen exposure results in a drastic change of the  Bi $4f$ core level spectrum where the narrow metallic Bi $4f$ doublet (with the  $4f$$_{7/2}$ component at a binding energy of 156.94~eV and the  $4f$$_{5/2}$ component obscured by the S $2p$$_{3/2}$ peak) is converted into a broad doublet with the $4f$$_{7/2}$ peak at  $159.05$~eV and the $4f$$_{5/2}$ component on the high binding energy side of the S $2p$$_{1/2}$ peak. A more efficient oxidation can be achieved by annealing the sample in air. The result of this procedure is shown by the dark blue lines in Fig.~\ref{fig:XPS_S2p}. These results demonstrate that all Bi is oxidized. Moreover, the binding energy of the Bi $4f$ peaks gives valuable information regarding the Bi oxide polymorph. It is consistent with the $\beta$-Bi$_2$O$_3$ phase \cite{Morgan:1973,Dharmadhikari:1982,Liu:2011ag,Chen:2021ab, Tian:2021}, which is known to be stabilized in thin-film and interface-supported geometries.

The corresponding XPS spectra in the Mo $3d$ and S $2s$ region are shown in Fig.~\ref{fig:XPS_S2p}(b). Bi intercalation merely leads to a shift of all the peaks to higher binding energy, consistent with the doping effect seen in the other core levels and the valence band. Upon oxidation the shift increases by about 80~meV. For strong oxidation in air (dark blue lines), a shoulder appears on the high binding energy side of the Mo $3d$ peaks.
This is interpreted as a minor contribution from Mo–O bonding, indicating limited oxidation of the MoS$_2$ layer under the most aggressive oxidation conditions.

The XPS results together with the preservation of the photoemission intensity related to the TMDC demonstrate that Bi is efficiently intercalated beneath SL MoS$_2$ and can be fully oxidized without destroying the SL MoS$_2$. The oxidation-induced changes are dominated by the transformation of the Bi layer into a $\beta$-Bi$_2$O$_3$-like phase, while only minor chemical modifications of the MoS$_2$ are observed. These findings support the interpretation that the pronounced electron accumulation observed by ARPES originates from a contact-mediated charge-transfer mechanism driven by the altered electronic properties of the oxidized Bi interlayer, rather than from extensive defect formation in the MoS$_2$ itself.

 \begin{figure*}
    \includegraphics[width=1\textwidth]{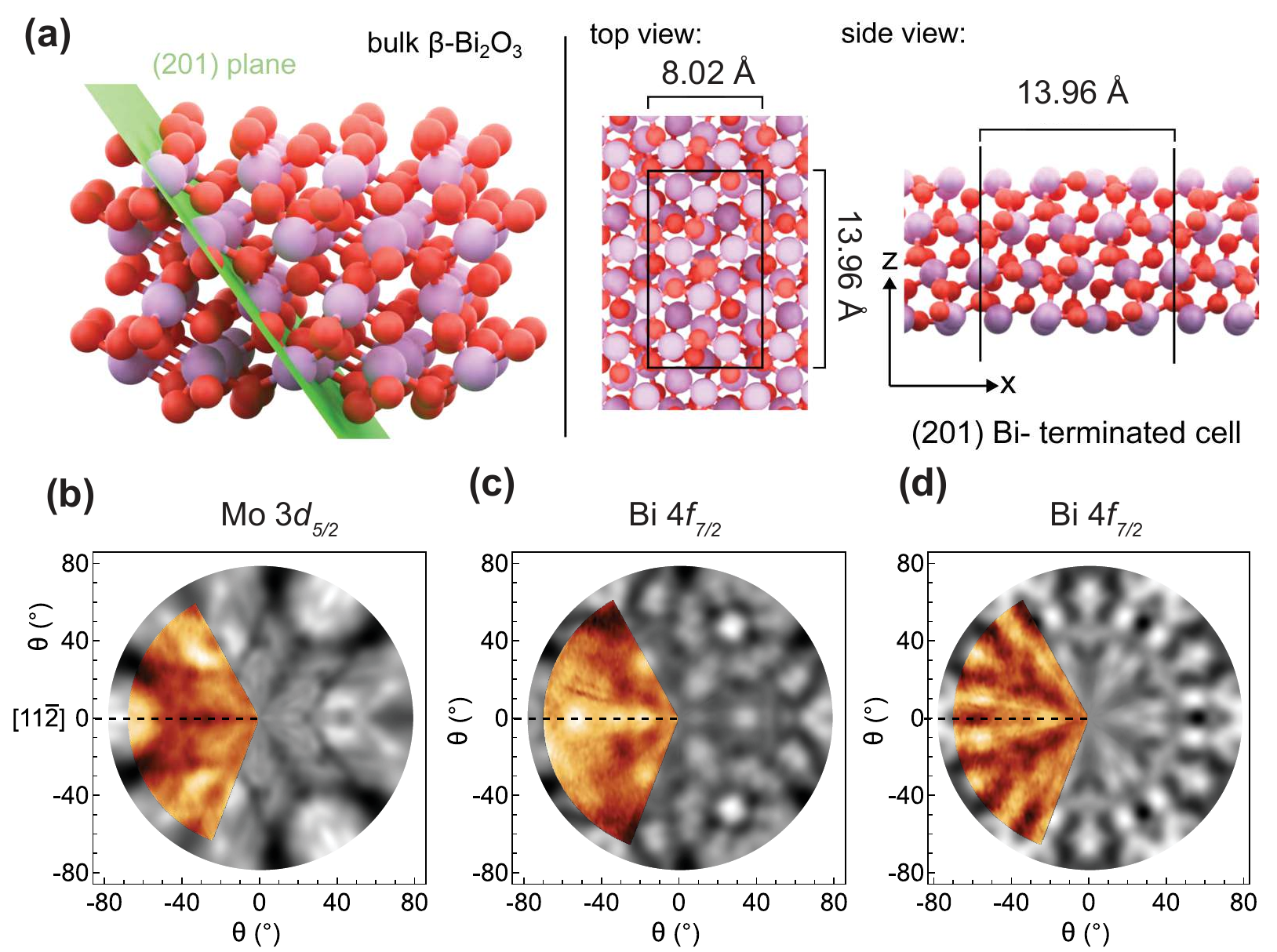}
     \caption{XPD characterization of the interface. (a) Structural model for the $\beta$-Bi$_2$O$_3$(201) surface assumed to be placed under SL MoS$_2$ in the Bi 4f$_{7/2}$ multiple scattering simulations. Stereographic projections of the modulation function  for (b) Mo $3d_{5/2}$ on $\beta$-Bi$_2$O$_3$ along with a simulation for  free-standing SL MoS$_2$, (c)  Bi $4f_{7/2}$ for $\beta$-Bi$_2$O$_3$ below SL MoS$_2$ with a simulation for  $\beta$-Bi$_2$O$_3$ without the  SL MoS$_2$ and (d) Bi $4f_{7/2}$ for $\beta$-Bi$_2$O$_3$ on Au(111) without SL MoS$_2$, along with a corresponding simulation. Note that the simulations in panels (c) and (d) differ by the number or rotational domains (3 \emph{vs.} 6).} 
     \label{fig:Diffraction}
\end{figure*}

Further structural information is obtained by X-ray photoelectron diffraction, \emph{i.e.}, by analyzing the angular distribution of the core level photoemission intensity \cite{Woodruff:2002aa,Bana:2018aa}. In this approach, the structure is determined by comparing an experimentally measured diffraction pattern to a multiple-scattering simulation for a particular trial structure, varying the structure until agreement with the experiment is reached.
In our case, the relevant best-fit substrate structure is the (201) surface of $\beta$-Bi$_2$O$_3$, used here as a representative local structural motif for the oxidized Bi interlayer rather than as a claim of long-range crystalline order. A structural model is given in  Fig.~\ref{fig:Diffraction}(a). This is also the surface for which we have calculated the work function by DFT. Fig.~\ref{fig:Diffraction}(b) shows the XPD modulation function, \emph{i.e.}, only the modulating part of the total photoemission intensity, for the Mo 3d$_{5/2}$ core level obtained at a photon energy of 360~eV after the intercalation and oxidation of bismuth. This is plotted together with a multiple-scattering simulation for free-standing SL MoS$_2$ with a single orientation \cite{Bana:2018aa}. It confirms that the Mo atoms are in the structural environment expected for MoS$_2$ and that the structural integrity is not compromised by the $\beta$-Bi$_2$O$_3$  intercalation. Fig.~\ref{fig:Diffraction}(c) probes the structure of the substrate via the Bi 4f$_{7/2}$ diffraction pattern from the bismuth oxide \emph{under} the SL MoS$_2$, acquired at a photon energy of h$\nu=650$~eV. The pattern agrees well with the structural model for  $\beta$-Bi$_2$O$_3$ when three rotational domains of the oxide in the model are accounted for. Note that the SL MoS$_2$ is not included in the simulation because it forms an incommensurate structure with the $\beta$-Bi$_2$O$_3$ substrate and its effect is thus averaged out. 
We thus conclude that the SL MoS$_2$ stays intact during the intercalation and oxidation and that the substrate can be well described by the  $\beta$-Bi$_2$O$_3$ structure. 
As a final confirmation, we present the corresponding Bi 4f$_{7/2}$ diffraction for $\beta$-Bi$_2$O$_3$ directly grown on Au(111) without SL MoS$_2$, acquired at h$\nu=650$~eV. We also find excellent agreement with the $\beta$-Bi$_2$O$_3$ structure, but only when assuming the presence of additional rotational domains (for details see Ref. \cite{Turoldo:2026aa}). Interestingly, the presence of the SL MoS$_2$ when oxidizing the intercalated Bi layer therefore restricts the number of oxide domains formed.  
These XPD results provide direct structural justification for using the $\beta$-Bi$_2$O$_3$-based model in the interpretation of the work function changes and the resulting charge-transfer doping.

\section{Conclusion}

In conclusion, we have demonstrated that controlled oxidation of an intercalated Bi contact layer leads to degenerate $n$-type doping of SL MoS$_2$, with carrier densities on the order of $10^{13}$~cm$^{-2}$ and a metallic electronic character in the contact region. In contrast, pristine Bi intercalation alone does not induce conduction-band occupation, underscoring the decisive role of interfacial chemistry in enabling charge transfer. 

These results establish oxidized Bi as an efficient $n$-type charge-transfer contact, conceptually complementary to the $\alpha$-RuCl$_3$-based contacts used to achieve degenerate $p$-type doping \cite{Pack:2024aa}. More generally, they suggest that interfacial oxidation can be exploited as a design strategy for contact engineering in two-dimensional semiconductors, enabling strong carrier doping without direct chemical modification of the 2D layer itself. The approach could possibly be combined with other nano-confinement growth methods \cite{Bian:2026}.

The absence of degenerate doping for SL MoS$_2$ on pristine Bi highlights that favorable band alignment is not guaranteed by the use of semimetal contacts alone, but depends sensitively on the detailed atomic and electronic structure of the interface. This emphasizes the importance of interface-specific control in the design of low-resistance contacts to 2D semiconductors.

\section{Methods}

Experiments were performed at the SGM-3 beamline of ASTRID2 \cite{Hoffmann:2004aa} and at the SuperESCA beamline of Elettra \cite{Abrami:1995aa}. At the SGM-3 beamline, ARPES and XPS measurements were performed at photon energies of h$\nu=54$~eV and 130~eV and an energy resolution better than 15~meV and 50meV, respectively. The angular resolution was better than 0.2~$^{\circ}$ and the sample  temperature was 35~K. STM measurements were carried out at room temperature employing a home built Aarhus-type microscope \cite{Laegsgaard:2001aa}.  XPS and XPD measurements at higher energies were carried out at the  SuperESCA beamline at the photon energies given in the text. All XPS binding energy scales refer to the measured Fermi level as zero. The sample was kept at room temperature for these measurements. Consistency of both preparations was ensured by comparing core level spectra and LEED patterns.

SL MoS$_2$ on Au(111) was prepared in ultra high vacuum following the methods presented in Refs. \cite{Miwa:2015ab,Gronborg:2015aa,Bana:2018aa}: 
The Au(111) substrate was mounted on a Ta holder and the temperature was measured by a thermocouple in tight contact with the crystal. A clean surface was obtained by repeated cycles of sputtering and annealing to 923~K. An atomically clean surface was confirmed by STM, showing the characteristic herringbone reconstruction, as well as the corresponding LEED pattern \cite{Hove:1981aa,Barth:1990aa}. In photoemission, the expected Au $4f$ surface core level shift was found \cite{Heimann:1981aa}, as well as the Rashba-split surface state \cite{LaShell:1996aa,Dendzik:2016aa}. 
The clean surface was then exposed at 823~K  to a sulphur and Mo flux coming from a home made Ag$\backslash$AgI$\backslash$Ag$_2$S electrochemical cell \cite{Heegemann:1975aa} and an e-beam evaporator or Mo filament, respectively. The combined deposition rate was about 0.057~monolayers per minute. 

Bi was intercalated under SL MoS$_2$ grown in this way by exposing the sample to a Bi flux from a heated crucible made of a Ta dispenser while keeping its temperature at 373~K. Finally, exposing the surface to $3 \times  10^{-5}$~mbar of O$_2$ through a doser very close to the sample (enabling a local pressure of at least $10^{-3}$~mbar \cite{Ulstrup:2014ad}) for 1~h at 423~K resulted in the oxidation of the Bi layer underneath the MoS$_2$ in a similar way as previously demonstrated for other intercalation systems \cite{Larciprete:2012aa}. More effective oxidation could be obtained by annealing the sample in air at 500~K for 20~min (see dark blue curves in Fig.~\ref{fig:XPS_S2p}). The bismuth oxide film without MoS$_2$ used for the control experiment in Fig.~\ref{fig:Diffraction}(d) and for the work function measurements was prepared in the same way, by first dosing Bi and then oxidizing the film in front of the doser, using the same conditions as with the MoS$_2$ present. The thickness of the Bi and $\beta$-Bi$_2$O$_3$ films were estimated by the attenuation of the Au 4$f$ core level peaks in XPS. 

XPD modulation functions were simulated using the program package Electron Diffraction in Atomic Clusters (EDAC) \cite{Garcia:2001}. The surfaces of the oxide were represented by Bi-terminated clusters using the experimentally determined bulk atomic positions  for $\beta$-Bi$_2$O$_3$ \cite{PerezMezcua:2016aa} and the calculated ones for the other polymorphs \cite{COD:2011}. 

Work function measurements were performed using photoemission by finding the position of the Fermi level edge $E_\mathrm{F}^{\mathrm{kin}}$ and the inelastic tail cutoff $E_\mathrm{tail}^{\mathrm{kin}}$ and then taking $\Phi = h\nu - \left( E_\mathrm{F}^{\mathrm{kin}} - E_\mathrm{tail}^{\mathrm{kin}} \right)$. For these measurements a small bias voltage was applied to the sample. 

DFT calculations were performed with Quantum Espresso \cite{Giannozzi_JoP2009, Giannozzi_JPCM2017} to compute the$\beta$-Bi$_2$O$_3$ work function, which was found to be 3.38 and 5.24~eV for the Bi and O-terminated bulk, respectively. This value for the Bi termination has been calculated before and our result is in good agreement with these earlier ones \cite{Tian:2021,Jing_ACSAMI2021}. 
Details of the structural relaxations, surface slab models, and work function extraction procedures are provided in the Appendix.

\section*{Acknowledgements}
Financial support by the Independent Research Fund Denmark  (Grants No. 1026-00089B and 4258-00002B) and by the Novo Nordisk Foundation (Grant number NFF23OC0085585) is gratefully acknowledged. A.T. acknowledges support from Collegio Universitario ``Luciano Fonda".

\section*{Appendix}

In this Appendix, we provide additional information on the gradual intercalation of Bi as observed by ARPES and the structural determination by low energy electron diffraction (LEED), scanning tunnelling microscopy (STM) and X-ray Photoelectron Diffraction (XPD), providing simulations obtained for other polymorphs of bismuth oxide and surface terminations of Bi$_2$O$_3$. We also discuss the experimental and theoretical determination of the work function.

\subsection{Bi intercalation in ARPES}

\begin{figure}
    \includegraphics[width=0.5\textwidth]{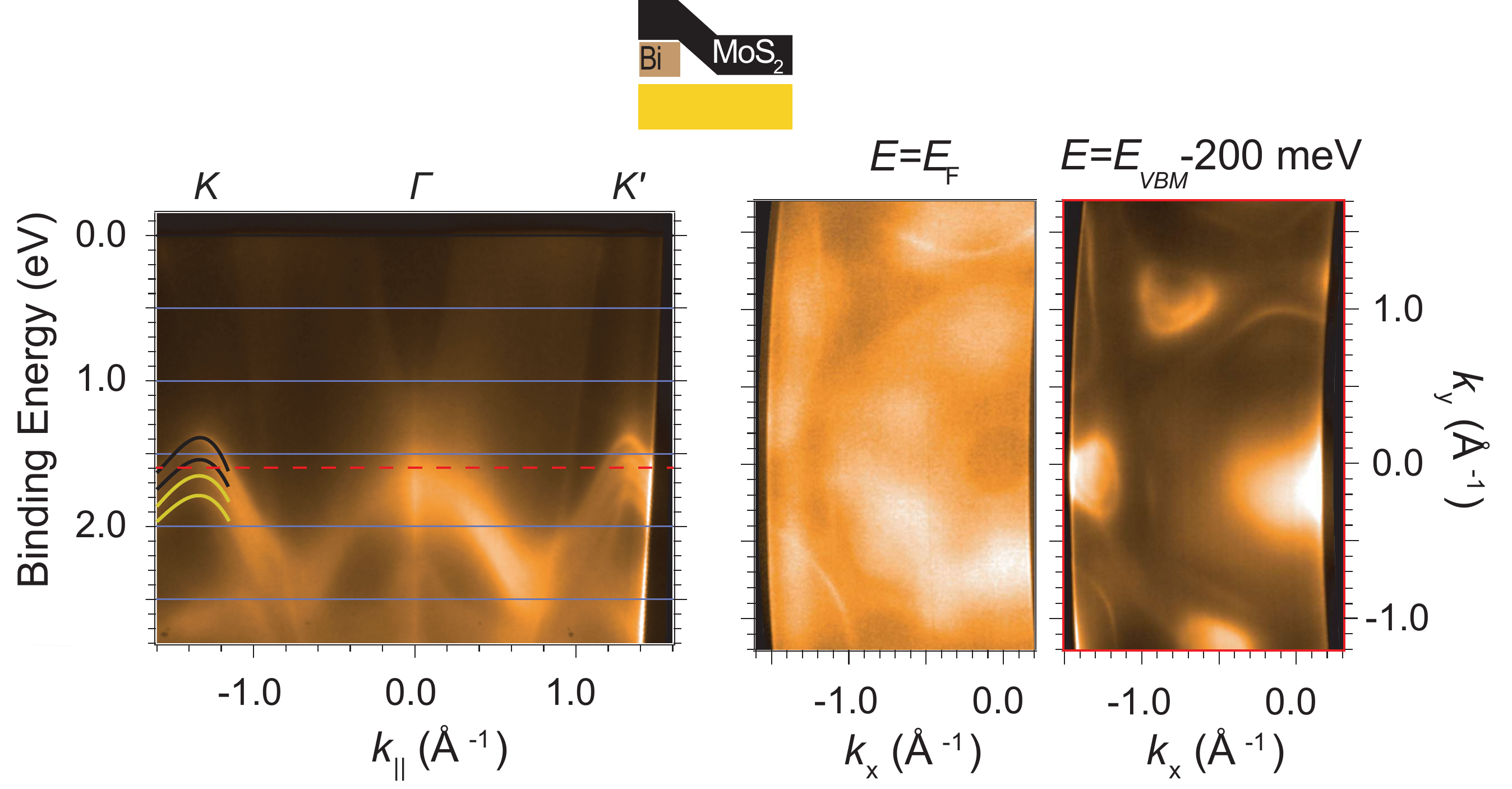}
     \caption{Bi intercalation at an intermediate state observed by ARPES. The panels correspond to those in Fig.~1 of the main paper but for partially intercalated single layer MoS$_2$. The original bands and shifted bands are marked by black and yellow curved lines, respectively. }
     \label{fig:S0}
\end{figure}

A partial intercalation of Bi, \emph{i. e.}, the situation between panels (a) and (b) of Fig.~\ref{fig:ARPES} in the main paper, is illustrated in Fig.~\ref{fig:S0}. At this stage, both the original single layer MoS$_2$ bands and the bands from the intercalated areas, shifted by 250~meV to higher binding energies, are visible. 

\subsection{Structural determination using scanning tunnelling microscopy and low energy electron diffraction}

\begin{figure*}
    \includegraphics[width=1\textwidth]{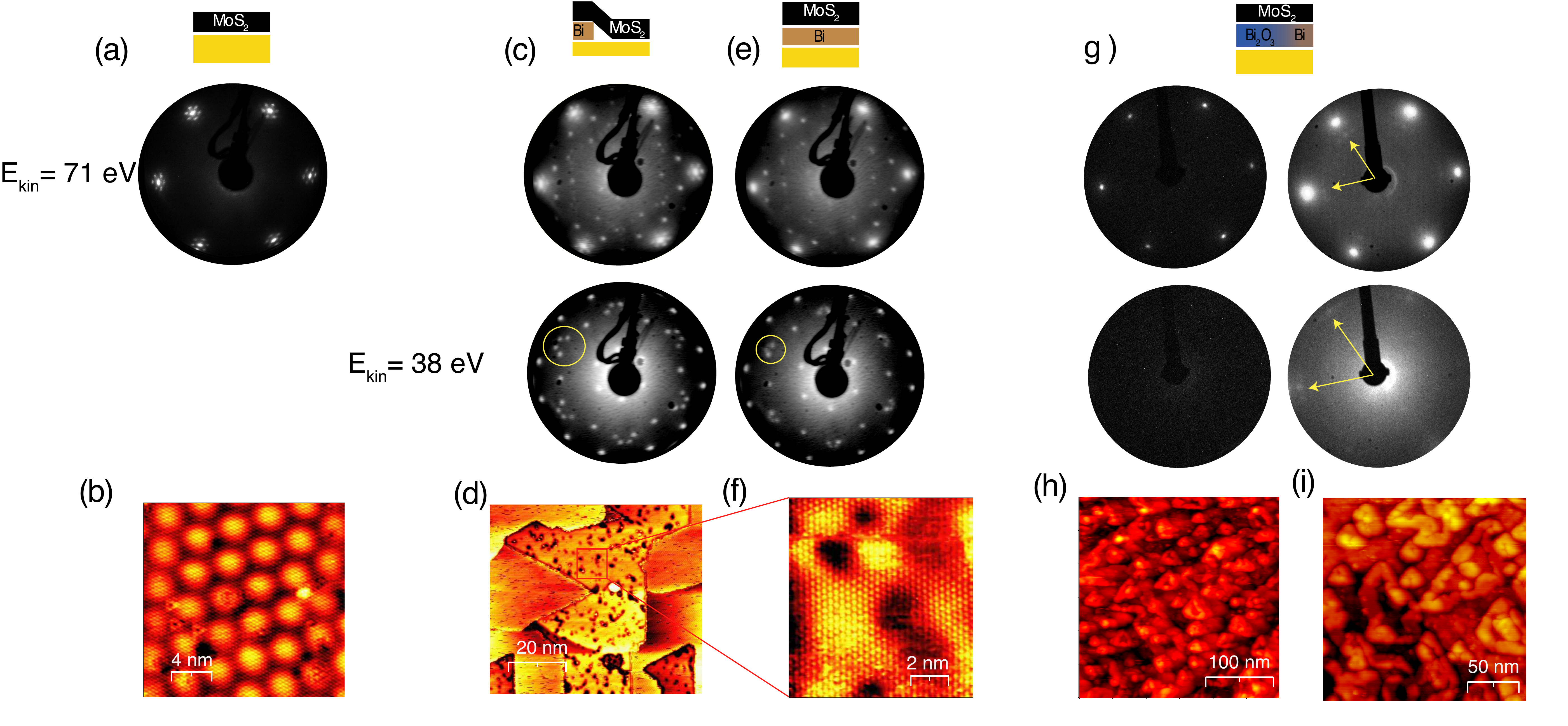}
     \caption{Bi intercalation and oxidation tracked by LEED and STM, following the same steps as shown in Fig.~1 of the main paper. (a), (b) Results for pristine SL MoS$_2$ on Au(111). (c), (d) Partially and (e), (f) fully Bi-intercalated situation. (g), (h) Corresponding data for SL MoS$_2$ on bismuth oxide. Scanning parameters: (b): $I_t=350$~pA, $V_t=111$~mV, (d) $I_t=260$~pA, $V_t=507$~mV, (f) $I_t=280$~pA, $V_t=730$~mV, (h): $I_t=360$~pA, $V_t=1044$~mV, (i): $I_t=730$~pA, $V_t=1022$~mV.}
     \label{fig:LEED}
\end{figure*}

The Bi intercalation and oxidation described in the main paper has also been tracked by LEED and STM. This is illustrated in Fig.~\ref{fig:LEED}. The results for SL MoS$_2$ on clean Au(111) in Fig.~\ref{fig:LEED}(a) and (b) show the characteristic interplay between the two slightly different lattice constants, resulting in a moir\'e pattern. In the LEED image, this is seen as the star-shaped pattern surrounding the substrate diffraction peaks. In STM, two hexagonal patterns are visible. The small-scale pattern corresponds to the atomic lattice of MoS$_2$ and the larger one is the moir\'e superstructure \cite{Miwa:2015aa,Bruix:2016aa}. 

The corresponding results for intermediate and full Bi intercalation are given in Fig.~\ref{fig:LEED}(c) - (f). The evolution of the LEED patterns shows two main trends: the loss of the moir\'e-related superstructure spots and the appearance of additional diffraction spots, signalling a new periodicity. 
Note that both LEED and ARPES average over macroscopic areas (approximately $1\times1$~mm$^2$ and $200\times100$$~\mu$m$^2$, respectively) that always contain contributions of different phases, in contrast to STM. Both the LEED and ARPES intensities can thus be understood as  incoherent superposition of intensity from all present phases. 
Here, the non-intercalated areas give rise to the moir\'e lattice spots. These are absent for the intercalated areas  due to the lack of direct SL MoS$_2$-Au(111) interaction and a gradual loss of the moir\'e spot intensity is expected, as the relative area of Bi-intercalated SL MoS$_2$ increases. The complicated pattern of additional diffraction spots upon Bi intercalation corresponds to that observed for depositing Bi directly on Au(111) \cite{He:2019aa,Chen:1993vg,Kawakami:2017ul,Vincente:2024,Turoldo:2026aa}. Especially the subtle change  in the diffraction pattern highlighted by the yellow circle at $E_\mathrm{kin} = 38$~eV is useful because it signals the transition from a $\sqrt{37}\times\sqrt{37}$~R$25.3^\circ$ reconstruction to a $p\times\sqrt{3}$ reconstruction that takes place when the Bi coverage exceeds 0.8~monolayer  \cite{He:2019aa,Kawakami:2017ul}. In STM, the intercalated SL MoS$_2$ can be identified by showing a hexagonal lattice with the MoS$_2$ periodicity on a background with some spatial variation but without the distinct moir\'e superstructure (see Fig.~\ref{fig:LEED}(d),~(f)). STM also reveals areas not covered by SL MoS$_2$ that show the expected Bi-induced reconstruction on Au(111) (not shown). Note that these results prove that intercalation takes place, but they do not give much information about structure of the intercalated Bi layer other than that it is either similar to that on clean Au(111), and thus gives the same diffraction pattern (or a pattern with a sub-set of the diffraction spots), or it is so disordered that it does not give any diffraction pattern. 

Finally, Fig.~\ref{fig:LEED}(g) and (h),(i) show LEED and STM data after oxidising the sample. At first sight the only diffraction spots that remain are those of SL MoS$_2$.  However, much weaker diffraction spots (marked by arrows in right images in panel (g)) can be observed when saturating the main diffraction spots, pointing towards some ordering of the oxide phase below. STM is still showing the characteristic triangular islands of the MoS$_2$. These appear partially lifted with protrusions near the edges, indicating a non-uniform formation of the oxide and an apparent corrugation that reaches about 20~\AA. The  formation of Bi oxide on Au(111) will be described in more detail in Ref. \cite{Turoldo:2026aa}. 

\subsection{XPD simulations of alternative structures}

\begin{figure*}
	\includegraphics[width=0.9\textwidth]{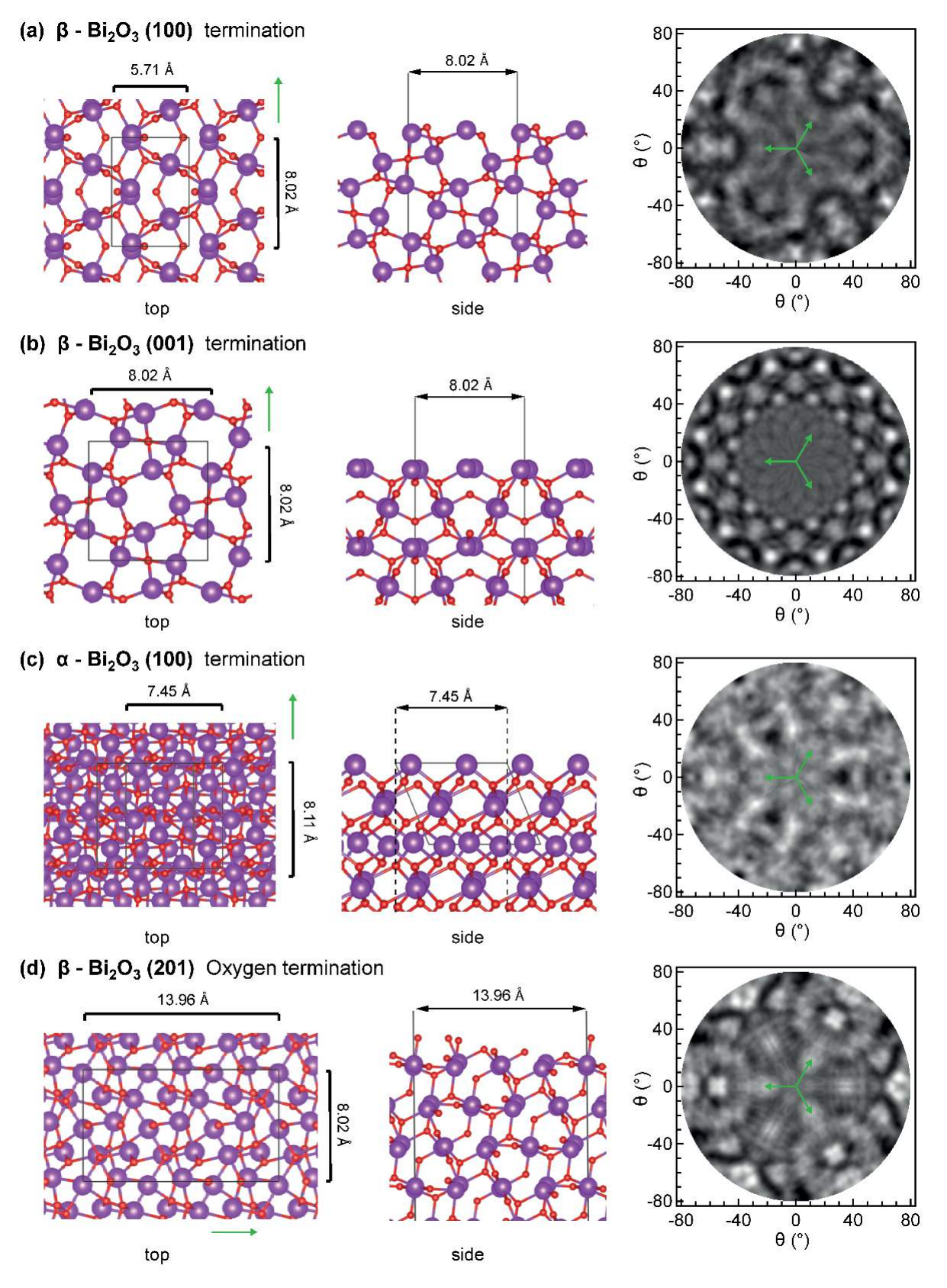}
	\caption{Alternatives to the Bi-terminated $\beta$-Bi$_2$O$_3$(201) structural model displayed in Fig.~3(c) of the main paper. XPD simulations for the Bi 4f$_{7/2}$ core level at $E_{kin} = 491$~eV, including three domains separated by $\pm120^\circ$ rotations (indicated by green arrows) for (a)  $\beta$-Bi$_2$O$_3$(100), (b)  $\beta$-Bi$_2$O$_3$(001), (c)  $\alpha$-Bi$_2$O$_3$(100), (d)  oxygen-terminated $\beta$-Bi$_2$O$_3$(201). }
	\label{fig:XPDsim}
\end{figure*}

Fig.~\ref{fig:Diffraction} shows excellent agreement between the measured XPD modulation functions and the calculations for the bismuth-terminated $\beta$-Bi$_2$O$_3$(201) surface, leading us to the conclusion that this represents the structure present at the interface. In order to exclude other Bi$_2$O$_3$ polymorphs and surface terminations, we performed the corresponding calculations for these as well. We considered only those structures compatible with the periodicity measured by LEED. The simulations are shown in Fig.~\ref{fig:XPDsim}. The locations of the high-intensity features in the stereographic projections of these simulated modulation functions do not match the experimental pattern, even when accounting for the possible coexistence of differently oriented domains with respect to the substrate. Interestingly, the simulation of the oxygen-terminated $\beta$-Bi$_2$O$_3$(201) surface is comparable—if not identical—to that of the bismuth-terminated one shown in  Fig.~\ref{fig:Diffraction}. This is not surprising because bismuth, being a heavy atom, has a much higher electron scattering cross section than oxygen and the additional oxygen atom do thus only have a minor effect on the resulting diffraction pattern. The oxygen termination can, however, be ruled out because of its much higher calculated work function of 5.24~eV (see theory section).

\subsection{Work function measurements}
\begin{figure}
	\includegraphics[width=0.4\textwidth]{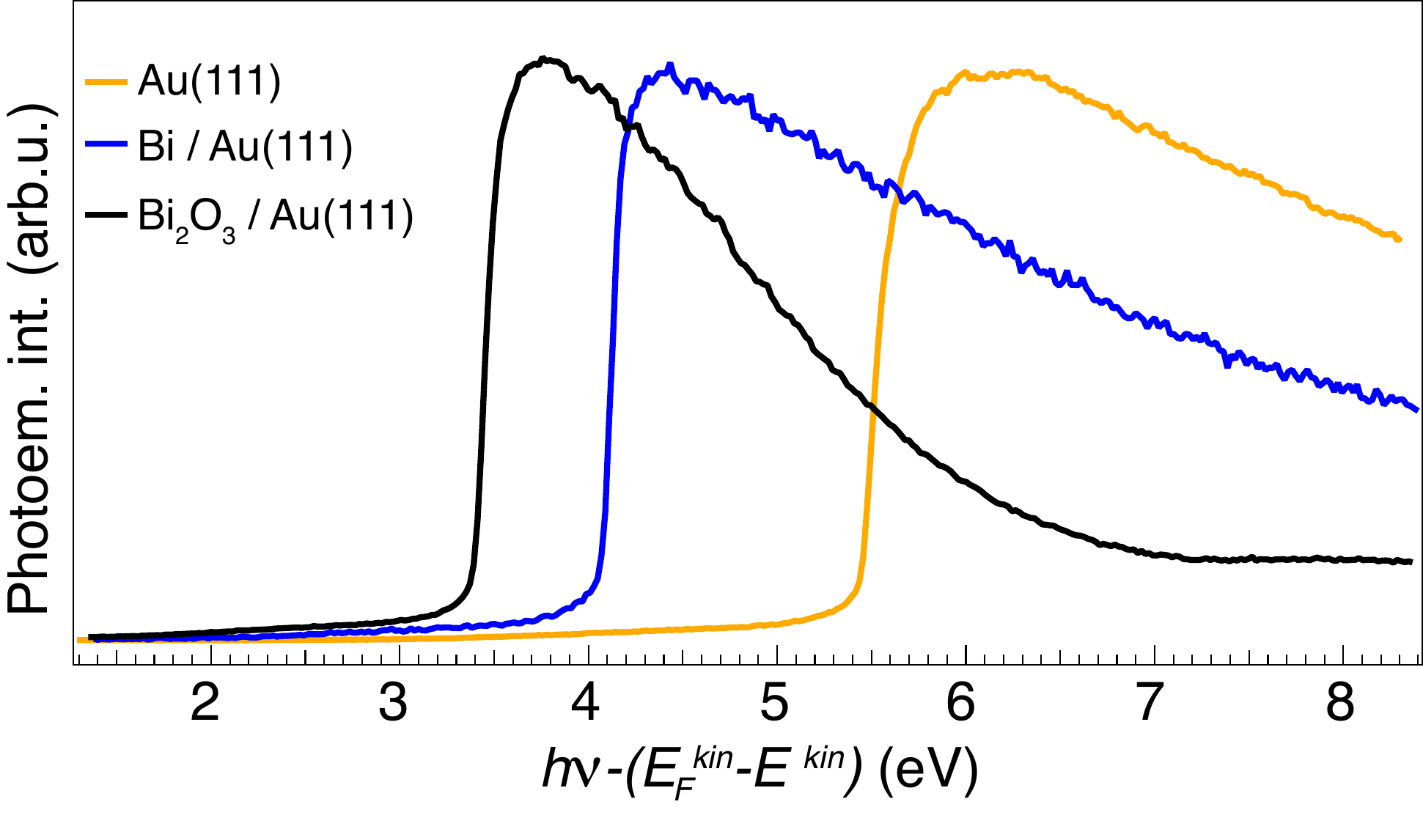}
	\caption{Photoemission intensity in the region of the secondary-electron as a function of  $h\nu~-~\left(~E_\mathrm{F}^{\mathrm{kin}} - E^{\mathrm{kin}} \right)$.}
	\label{fig:Wf1}
\end{figure}

This section presents the spectra acquired to determine the work function of each surface preparation as discussed in the main text.
Figure~\ref{fig:Wf1} shows the photoemission intensity in the secondary-electrons region for the clean Au(111), Bi/Au(111) and  Bi$_2$O$_3$/Au(111). The data are plotted on a scaled energy axis defined as $h\nu~-~\left(~E_\mathrm{F}^{\mathrm{kin}} - E^{\mathrm{kin}} \right)$, such that the work function can be directly read from the middle of the steep low-energy cut-off.

\subsection{First-Principles Calculations of the Work Function}

\begin{figure}[h]
    \includegraphics[width=0.4\textwidth]{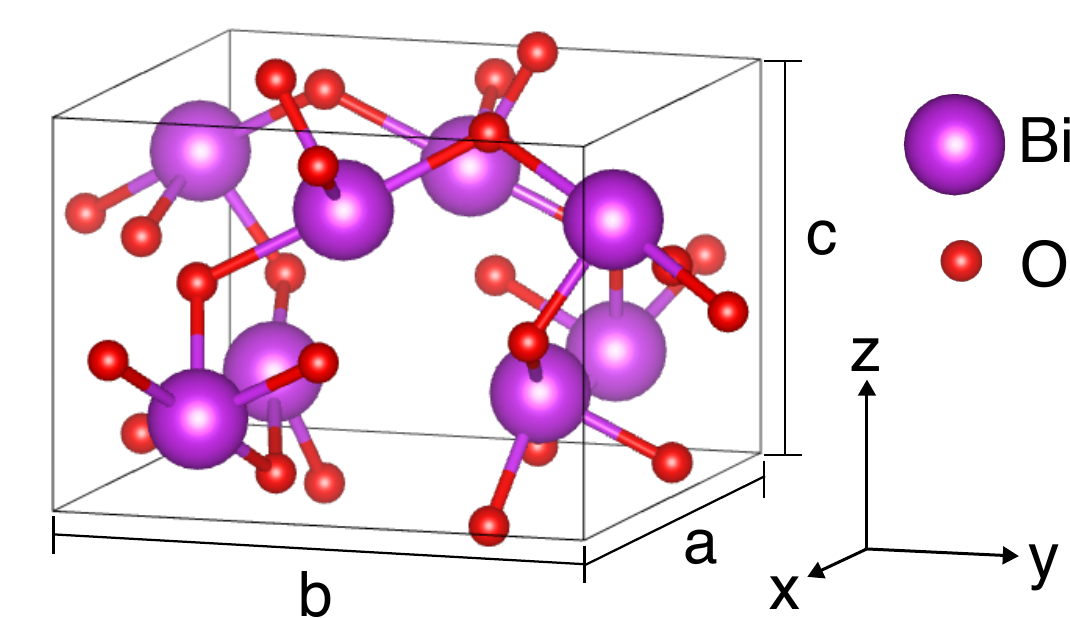}
     \caption{Unit cell of relaxed $\beta$-Bi$_2$O$_3$. Lattice parameters are a~=~8.01996~\AA, b~=~8.01996~\AA, c~=~5.7113~\AA.}
     \label{fig:Bi2O3bulk}
\end{figure}

Density functional theory (DFT) calculations were performed using Quantum Espresso \cite{Giannozzi_JoP2009, Giannozzi_JPCM2017} to determine the work function of the $(201)$ surface of $\beta$-Bi$_2$O$_3$.
We employed  fully relativistic Projector Augmented-Wave pseudopotentials \cite{Dal-Corso:2014aa}, consistent with the Perdew–Burke–Ernzerhof exchange–correlation functional, and included spin–orbit coupling. The kinetic energy cutoff was set to 60 Ry ($\sim$816~eV) and a $6$$\times$$6$$\times$$8$ Monkhorst-Pack ${\bf k}$-point grid was used to sample the Brillouin zone. 
The bulk Bi$_2$O$_3$ structure was first fully relaxed to determine its equilibrium lattice parameters as shown in Fig.~\ref{fig:Bi2O3bulk}

\begin{figure}
    \includegraphics[width=0.4\textwidth]{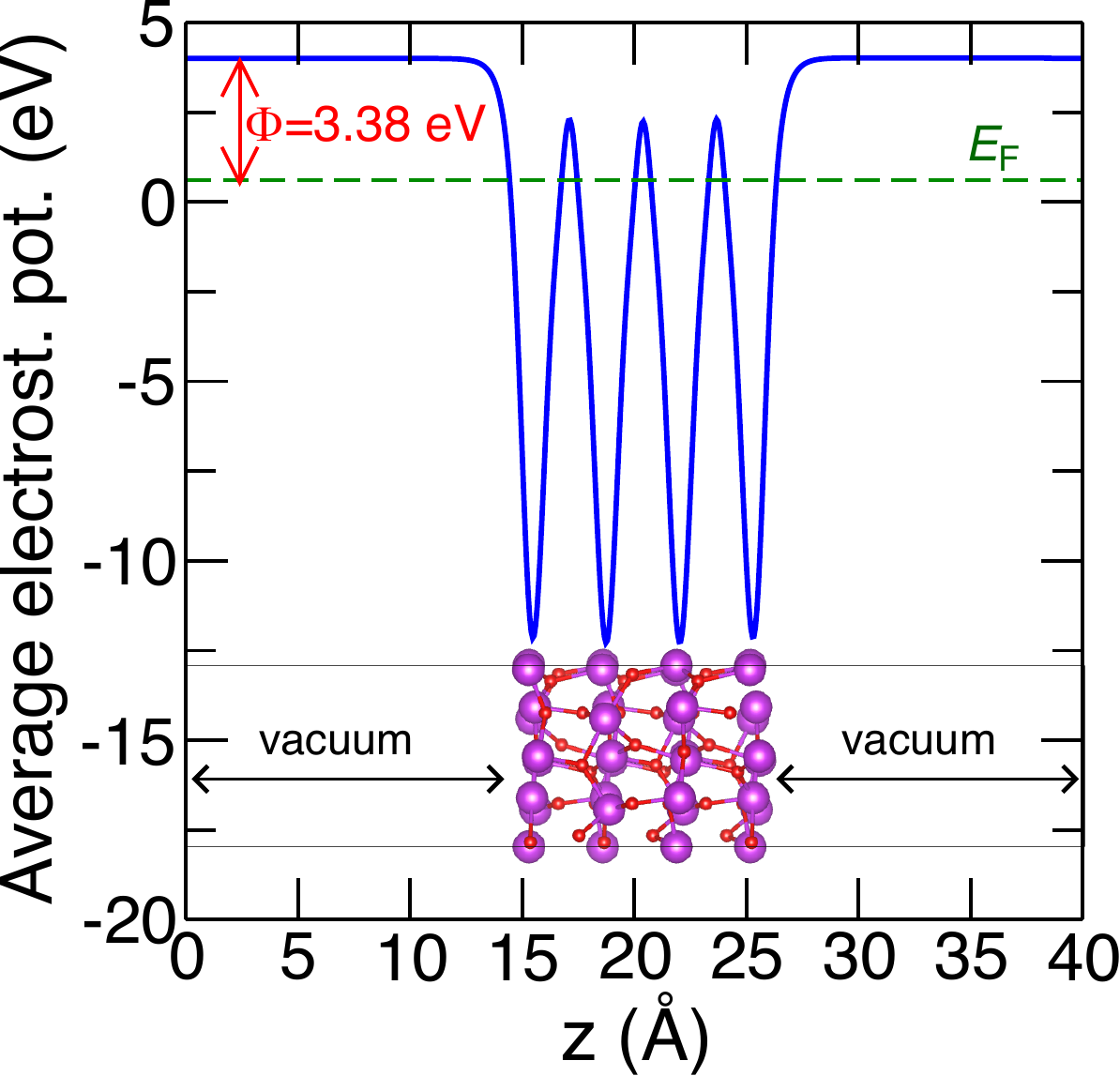}
     \caption{Plane-averaged electrostatic potential energy along the $z$ direction for the Bi-terminated (201) Bi$_2$O$_3$ slab. $E_\mathrm{F}$ is the Fermi energy. }
     \label{fig:AveragePotential}
\end{figure}

A (201) surface slab was then constructed, and both Bi- and O-terminated surfaces were considered for subsequent analysis. 
Given that Quantum Espresso assumes periodic boundary conditions in all spatial directions, to prevent spurious interactions between periodic images of the slab along the $z$--axis, a vacuum region of 15 \AA\  was inserted above and below the slab and also dipole-correction corrections were applied \cite{Giannozzi_JPCM2017}. 
Finally, the plane-averaged electrostatic potential energy, which includes the Hartree and bare ionic contributions, was calculated. The result for the Bi-terminated surface is shown in Fig.~\ref{fig:AveragePotential}. From the alignment of the vacuum potential energy  with the Fermi energy $E_\mathrm{F}$, we obtained a work function of 3.38 eV, in good agreement with previously reported values in the literature \cite{Tian:2021,Jing_ACSAMI2021}. A corresponding calculation for the O-terminated surface leads to a high work function of 5.24~eV that cannot be reconciled with the experimental findings.

\clearpage


\end{document}